\documentclass[showpacs]{revtex4}
\usepackage{epsfig}
\def\beq{\begin{equation}}
\def\enq{\end{equation}}
\def\beqa{\begin{eqnarray}}
\def\enqa{\end{eqnarray}}

\def\MeV{\nobreak\,\mbox{MeV}}
\def\GeV{\nobreak\,\mbox{GeV}}

\def\qq{\lag\bar{q}q\rag}

\def\qqs{\lag\bar{s}s\rag}

\def\mixs{\lag\bar{s}g\si.Gs\rag}
\def\Gd{\lag g^2G^2\rag}
\def\G3{\lag g^3G^3\rag}

\def\rh{\rho}
\def\si{\sigma}

\def\al{\alpha}
\def\be{\beta}
\def\alma{\alpha_{max}}
\def\almi{\alpha_{min}}
\def\bemi{\beta_{min}}
\def\lb{\label}
\def\nn{\nonumber}

\newcommand{\rag}{\rangle}
\newcommand{\lag}{\langle}

\begin{document}

\title{\sc
QCD sum rules study of the $J^{PC}=1^{--}$ charmonium $Y$ mesons
}
\author{R.M. Albuquerque}
\email{rma@if.usp.br}
\affiliation{Instituto de F\'{\i}sica, Universidade de S\~{a}o Paulo,
C.P. 66318, 05389-970 S\~{a}o Paulo, SP, Brazil}
\author{M. Nielsen}
\email{mnielsen@if.usp.br}
\affiliation{Instituto de F\'{\i}sica, Universidade de S\~{a}o Paulo,
C.P. 66318, 05389-970 S\~{a}o Paulo, SP, Brazil}

\begin{abstract}
We use QCD  sum rules to test the nature of the recently observed mesons
$Y(4260)$, $Y(4350)$ and $Y(4660)$, assumed to be  exotic four-quark 
$(c\bar{c}q\bar{q})$ or $(c\bar{c}s\bar{s})$ states with $J^{PC}=1^{--}$. 
We work at leading order in $\alpha_s$, consider the contributions of higher
dimension condensates and keep terms which are linear in
the strange quark mass $m_s$. We find for the $(c\bar{c}s\bar{s})$ state
a mass $m_Y=(4.65\pm 0.10)$  GeV which is compatible 
with the experimental candidate $Y(4660)$, while  for the $(c\bar{c}q\bar{q})$
state we find a mass $m_Y=(4.49\pm 0.11)$  GeV, which is higger than the mass
of the experimental candidate $Y(4350)$. With the tetraquark structure we are
working we can not explain the $Y(4260)$ as a tetraquark state.  We also 
consider molecular $D_{s0}\bar{D}_s^*$ and $D_{0}\bar{D}^*$ states. For the
$D_{s0}\bar{D}_s^*$ molecular  state we get $m_{D_{s0}\bar{D}_s^*}=(4.42\pm 
0.10)$  GeV which is consistent, considering the errors, with the mass
of the meson $Y(4350)$ and for the $D_{0}\bar{D}^*$ molecular  state we get 
$m_{D_{0}\bar{D}^*}=(4.27\pm 0.10)$  GeV in excelent agreement with the mass
of the meson $Y(4260)$.

\end{abstract}

\maketitle

\section{Introduction}

Recent experimental observations of new particles at $B$-factories, have 
revitalized the field of hadron spectroscopy.  While some of these observed 
states, like the $X(3940)$ \cite{belle1}, can be understood as $c\bar{c}$ 
states \cite{vol}, these observations
suggest that the spectrum of the charmonium states is much more rich than
suggested by the quark-antiquark model and may include tetraquark or molecular
states. In particular, the $X(3872)$ \cite{belle2} , with quantum numbers 
$J^{PC} = 1^{++}$, was the first observed state that did not fit easily  the 
charmonium spectrum. It also presents a strong isospin violating decay that 
disfavors a $c \bar{c}$ assignment. The $Z^+(4430)$, recently observed in the 
$Z^+\to \psi^\prime\pi^+$ decay mode
\cite{belle3}, is the most intriguing one since, being a charged state, it
can not be a pure $c\bar{c}$ state.

Besides the $X(3872)$ and $Z^+(4430)$ mesons, at least three peaks of the 
$J^{PC}=1^{--}$ family: the $Y(4260)$ \cite{babar1}, the $Y(4325)$ 
\cite{babar2} or/and $Y(4360)$ \cite{belle4} (that here we call $Y(4350)$) 
and the $Y(4660)$ \cite{belle4} can not be 
easily fited in the charmonium spectrum \cite{zhu}. Since the only 
observed decay channels
for these states are those containing the $J/\psi$ (for $Y(4260)$) or $\psi'$
(for the others) plus a pair of pions, in ref.~\cite{vol} this was considered 
as  an indication that these mesons contain a particular charmonium resonance,
$J/\psi$ or $\psi'$, that stays intact inside a more complex hadronic 
structure.

A critical information for understanding the structure of these states is 
wether
the pion pair comes from a resonance state. From the di-pion invariant mass 
spectra shown in ref.~\cite{babarconf} there is some indication that only the
$Y(4660)$ has a well defined intermediate state consistent with $f_0(980)$
\cite{babarconf}. Due to this fact and the proximity of the mass of the
$\psi'-f_0(980)$ system with the mass of the $Y(4660)$ state, in 
ref.~\cite{ghm},
the $Y(4660)$ was considered as a $f_0(980)~\psi'$ bound state. The $Y(4660)$
was also suggested to be a baryonium state \cite{qiao} and a canonical  
5 $^3$S$_1$ $c\bar{c}$ state \cite{dzy}.

In the case of $Y(4260)$, in ref.~\cite{maiani} it was considered as a 
$sc$-scalar-diquark $\bar{s}\bar{c}$-scalar-antidiquark in a $P$-wave state. 
In a naive estimate, the mass of a $sc$-scalar-diquark would be approximately 
equal to the mass of the $D_s$ meson. Since a unit of angular momentum can be 
estimated as the mass difference between the nucleon and the lowest-lying odd 
parity excited state, and it is about 600 MeV \cite{gerry}, one would expect 
the mass of a $sc$-scalar-diquark $\bar{s}\bar{c}$-scalar-antidiquark in a 
$P$-wave state about 4540 MeV, almost 300 MeV above the $Y(4260)$ mass.

Here we use the QCD sum rules (QCDSR) \cite{svz,rry,SNB} to study the 
two-point function  for a diquark-antidiquark tetraquark state
with a symmetric spin distribution: $[cs]_{S=1}[\bar{c}\bar{s}]_{S=0}+
[cs]_{S=0}[\bar{c}\bar{s}]_{S=1}$, to see if any of these new $Y$ mesons
can be described by such a current.
In  previous calculations, the QCDSR approach was used to study
the $X(3872)$ considered as a diquark-antidiquark state \cite{x3872},
and the $Z^+(4430)$ meson, considered as a $D^*D_1$ molecular state 
\cite{nos}.
In both cases a very good agreement with the experimental mass was obtained.

\section{The two-point correlator}

The lowest-dimension interpolating operator for describing  a $J^{PC}=1^{--}$ 
state with the symmetric spin distribution: $[cs]_{S=0}[\bar{c}\bar{s}]_{S=1}+
[cs]_{S=1}[\bar{c}\bar{s}]_{S=0}$ is given
by: 
\beq
j_\mu={\epsilon_{abc}\epsilon_{dec}\over\sqrt{2}}[(s_a^TC\gamma_5c_b)
(\bar{s}_d\gamma_\mu\gamma_5 C\bar{c}_e^T)+(s_a^TC\gamma_5\gamma_\mu c_b)
(\bar{s}_d\gamma_5C\bar{c}_e^T)]\;,
\label{field}
\enq

where $a,~b,~c,~...$ are color indices and $C$ is the charge conjugation
matrix. 

The two-point correlation function is given by:
\beq
\Pi_{\mu\nu}(q)=i\int d^4x ~e^{iq.x}\lag 0
|T[j_\mu(x)j^\dagger_\nu(0)]
|0\rag=-\Pi(q^2)(g_{\mu\nu}q^2-q_\mu q_\nu),
\lb{2po}
\enq
from where we get 
\beq
\Pi_{\mu}^{\mu}(q)=-3q^2\Pi(q^2),
\lb{inv}
\enq

The calculation of the phenomenological side at the 
hadron level proceeds by writing a dispersion relation for the invariant 
function in Eq.~(\ref{inv}):
\beq
\Pi^{phen}(q^2)=-\int ds\, {\rho(s)\over q^2-s+i\epsilon}\,+\,\cdots\,,
\label{phen}
\enq
where $\rho$ is the spectral density and the dots represent subtraction 
terms. The spectral density is described, as usual, as a single sharp
pole representing the lowest resonance plus a smooth continuum representing
higher mass states:
\beqa
\rho(s)&=&\lambda^2\delta(s-m_{Y}^2) +\rho^{cont}(s)\,,
\label{den}
\enqa
where $\lambda$ is proportional to the meson decay constant $f_Y$, which 
parametrizes the coupling of the vector meson, $Y$, to the current $j_\mu$:
\beq\label{eq: decay}
\lag 0 |
j_\mu|Y\rag =f_Ym_Y^4\epsilon_\mu~=~\lambda m_Y\epsilon_\mu~. 
\enq

For simplicity, it is
assumed that the continuum contribution to the spectral density,
$\rho^{cont}(s)$ in Eq.~(\ref{den}), vanishes bellow a certain continuum
threshold $s_0$. Above this threshold, it is assumed to be given by
the result obtained with the OPE. Therefore, one uses the ansatz \cite{io1}
\beq
\rho^{cont}(s)=\rho^{OPE}(s)\Theta(s-s_0)\;,
\enq
with
\beq
\rho^{OPE}(s)={1\over\pi}Im[\Pi^{OPE}(s)]\;.
\enq

In the OPE side, we work at leading order in $\alpha_s$ and consider the
contributions of condensates up to dimension six.  We keep the term
which is linear in the strange-quark mass $m_s$. We use the 
momentum space expression for the charm quark propagator,
while the light-quark part of the correlation function is
calculated in the coordinate-space. 
The correlation function, $\Pi(q^2)$, in the OPE side can also be written 
in terms of a dispersion relation:
\beq
\Pi^{OPE}(q^2)=\int_{4m_c^2}^\infty ds {\rho^{OPE}(s)\over s-q^2}\;.
\lb{ope}
\enq
Therefore, after making an inverse-Laplace (or Borel) transform on both 
sides, and
transferring the continuum contribution to the OPE side, the sum rule
for the  vector meson $Y$ can be written as
\beq \lambda^2e^{-m_Y^2/M^2}=\int_{4m_c^2}^{s_0}ds~
e^{-s/M^2}~\rho(s)\;, \lb{sr} \enq
where 
\beq
\rho^{OPE}(s)=\rho^{pert}(s)+\rh^{\qqs}(s)+\rh^{\lag G^2\rag}(s)
+\rh^{mix}(s)+\rh^{\qqs^2}(s)\;,
\lb{rhoeq}
\enq
with
\beqa\label{rhoope}
&&\rho^{pert}(s)=-{1\over 2^{8}3 \pi^6s}\int\limits_{\almi}^{\alma}
{d\al\over\alpha^3}
\int\limits_{\bemi}^{1-\al}{d\be\over\be^3}(1-\al-\be)
\left[(\al+\be)m_c^2-\al\be s\right]^3\left[m_c^2-2m_c^2(\alpha+\beta)+\al\be 
s\right],
\nn\\
&&\rho^{\qqs}(s)=-{3m_sm_c^2\qqs\over2^{4} \pi^4s}\int\limits_{\almi}^{\alma}
{d\al\over\alpha}
\int\limits_{\bemi}^{1-\al}{d\be\over\be}\left[(\al+\be)m_c^2-\al\be s\right],
\nn\\
&&\rho^{\lag G^2\rag}(s)={m_c^2\Gd\over3^22^{9}\pi^6s}
\int\limits_{\almi}^{\alma} {d\al}
\int\limits_{\bemi}^{1-\al}{d\be\over\beta^3}(1-\al-\be)\left[4(2\al+2\be-1)
m_c^2-{3m_c^2\beta\over\alpha}-\be s(7\alpha-3)\right],
\nn\\
&&\rho_1^{mix}(s)={m_c\mixs\over2^{6}3 \pi^4s}\int\limits_{\almi}^{\alma}
{d\al\over\alpha}
\int\limits_{\bemi}^{1-\al}{d\be\over\be^2}(2\alpha+\beta)\left[(\al+\be)m_c^2
-\al\be s\right],
\nn\\
&&\rho_2^{mix}(s)=-{m_s\mixs\over2^{6}3 \pi^4s}\int\limits_{\almi}^{\alma}
{d\al}\left[10m_c^2-2\alpha(1-\alpha)s-{m_c^2-\alpha(1-\alpha)s\over1-\alpha}
-5m_c^2\int\limits_{\bemi}^{1-\al}{d\be\over\be}\right],
\nn\\
&&\rho^{\qqs^2}(s)=-{\qqs^2\over 36\pi^2}\left({5m_c^2\over s}-{1\over2}
\right)\sqrt{1-4m_c^2/s},
\enqa

\beqa
&&\rho^{mix\qqs^2}(s)=-{\qqs\mixs\over3^22^4\pi^2s}(1+4m_c^2/s)
\sqrt{1-4m_c^2/s},\nn\\
&&\Pi^{mix\qqs}(M^2)=-{\qqs\mixs\over3^22^4 \pi^2}\left({2\over3}-3\int_0^1
d\al\,\exp\!\left[{-{m_c^2
\over\al(1-\al)M^2}}\right]\bigg[\al-2\alpha^2+{2m_c^2
\over M^2}\bigg]\right)\,.
\label{8m2}
\enqa
where the integration limits are given by $\almi=({1-\sqrt{1-
4m_c^2/s})/2}$, $\alma=({1+\sqrt{1-4m_c^2/s})/2}$ and $\bemi={\al
m_c^2/( s\al-m_c^2)}$.
The contribution of dimension-six condensates $\lag g^3 G^3\rag$
is neglected, since it is suppressed by   the loop factor $1/16\pi^2$.
In Eq.~(\ref{8m2}) we have included, for completeness a part of the 
dimension-8 condensate, related with the mixed condensate-quark condensate
contribution. We should note that a complete evaluation of these 
contributions 
require more involved analysis including a non-trivial choice
of the factorization assumption basis \cite{BAGAN}.
In Eq.~(\ref{8m2}) the $\Pi^{mix\qqs}(M^2)$ term is treated separately because
its imaginary part is proportional to delta functions and, therefore, can be 
easily integrated \cite{hen}.

It is very interesting to notice that the current in Eq.~(\ref{field}) does
not get contribution from the quark condensates when $m_s=0$. This is very 
different
from the OPE behavior obtained to the scalar-diquark axial-antidiquark current
used for the $X(3872)$ meson in ref.~\cite{x3872}, but very similar to the OPE
behavior obtained for the axial double-charmed meson $T_{cc}$, also
described by a scalar-diquark axial-antidiquark current \cite{tcc}.

\section{Sum rule  predictions for  $m_Y$} \unboldmath

The values used for the quark
masses and condensates are \cite{SNB,narpdg}:
$m_c(m_c)=(1.23\pm 0.05)\,\GeV $, $m_s=(0.13\pm 0.03)\,\GeV $,
$\lag\bar{q}q\rag=\,-(0.23\pm0.03)^3\,\GeV^3$, $\qqs=0.8\qq$,
$\lag\bar{q}g\si.Gq\rag=m_0^2\lag\bar{q}q\rag$ with $m_0^2=0.8\,\GeV^2$,
$\lag g^2G^2\rag=0.88~\GeV^4$.

We evaluate the sum rules in the Borel range $2.8 \leq M^2 \leq 4.6\GeV^2$,
and in the $s_0$ range $5.0\leq \sqrt{s_0} \leq5.2$ GeV. To fix the continuum
threshold range we extract the mass from the sum rule, for a given $s_0$,
and accept such value of $s_0$ if the obtained mass is around 0.5 GeV smaller
than $\sqrt{s_0}$.

\begin{figure}[h]
\centerline{\epsfig{figure=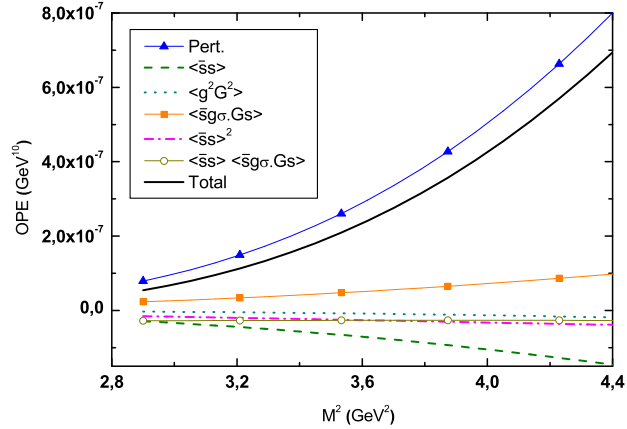,height=70mm}}
\caption{The OPE convergence in the region $2.8 \leq M^2 \leq
4.5~\GeV^2$ for $\sqrt{s_0} = 5.1$ GeV. Perturbative
contribution (solid line with triangles),  $\qqs$ contribution (dashed-line),
$\langle g^2G^2\rangle$ contribution (dotted line), $\mixs$ contribution 
(solid line with squares), $\langle \bar{s}s\rangle^2$ contribution 
(dot-dashed line), $\mixs\qqs$ contribution (solid line with spheres) 
and total contribution (solid line).}
\label{figconvtcc}
\end{figure}

From Fig.~\ref{figconvtcc} we see that we obtain a quite good OPE
convergence for $M^2\geq 3.2$ GeV$^2$. Therefore, we  fix the lower
value of $M^2$ in the sum rule window as $M^2_{min} = 3.2$ GeV$^2$.
This figure also shows that dimension-eight condensate contributions
is very small. Since this is not a complete evaluation of the dimension-eight
condensates contribution, it will be neglected in this calculation.

\begin{figure}[h]
\centerline{\epsfig{figure=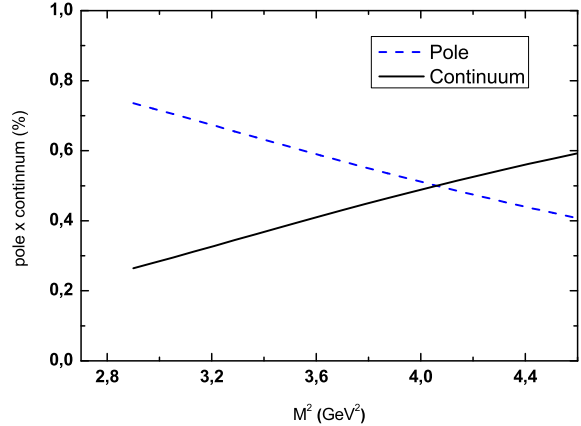,height=70mm}}
\caption{The dashed line shows the relative pole contribution (the
pole contribution divided by the total, pole plus continuum,
contribution) and the solid line shows the relative continuum
contribution for $\sqrt{s_0}=5.1~\GeV$.}
\label{figpvc}
\end{figure}

In Fig.~\ref{figpvc} we show the comparison between pole and
continuum contributions for $\sqrt{s_0} = 5.1$ GeV, and we see that
for $M^2\leq4.05~\GeV^2$, the pole contribution is bigger than the continuum
contribution. Therefore, we fix $M^2=4.05~\GeV^2$ as the upper limit of the 
Borel window for $\sqrt{s_0} = 5.1$ GeV.
The same analysis for the other values of the
continuum threshold gives $M^2 \leq 3.8$  GeV$^2$ for $\sqrt{s_0} = 5.0~\GeV$
and $M^2 \leq 4.2$  GeV$^2$ for $\sqrt{s_0} = 5.2~\GeV$.
We then consider, for each value of $s_0$, the range of $M^2$ values 
from 3.2 $\GeV^2$ until the one allowed by the pole dominance criteria given 
above.

\begin{figure}[h]
\centerline{\epsfig{figure=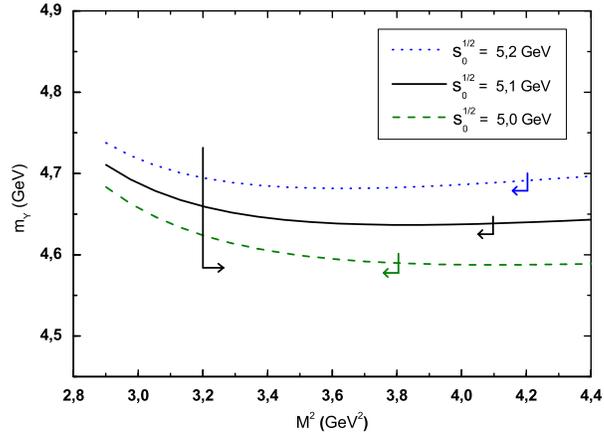,height=70mm}}
\caption{The $Y$ meson mass as a function of the sum rule parameter
($M^2$) for different values of $\sqrt{s_0}$. The arrows
indicate the region allowed for the sum rules: the lower limit
(cut below 3.2 GeV$^2$) is given by OPE convergence requirement and the
upper limit by the dominance of the QCD pole contribution.}
\label{figmx}
\end{figure}

To extract the mass $m_Y$ we take the derivative of Eq.~(\ref{sr})
with respect to $1/M^2$, and divide the result by Eq.~(\ref{sr}).
In Fig.~\ref{figmx}, we show the $Y$ meson mass, for different values of
$\sqrt{s_0}$, in the relevant sum rule window, with the upper and lower 
validity limits indicated.  From this figure we see that we get a very good 
Borel stability for $m_Y$.

To check the dependence of our results with the value of the
charm quark mass, we fix $\sqrt{s_0}=5.1~\GeV$ and vary the quark masses
and the quark condensate in the range given above. We see that the results
are not very sensitive to the changes in the values of the quark masses and 
condensate, the main source of uncertainty being the continuum threshold.
Adding the errors in quadrature we finally get
\beq
m_Y = (4.65\pm0.10)~\GeV,
\label{ymass}
\enq
in excelent agreement with the mass of the $Y(4660)$ meson. Therefore we 
conclude that the meson $Y(4660)$ can be described by a tetraquark state with
a spin configuration given by scalar and vector diquarks.

\section{Sum rule study for $Y(4350)$}
As pointed out in the introduction, from the di-pion invariant mass 
spectra shown in ref.~\cite{babarconf}, there is some indication that only the
$Y(4660)$ has a well defined intermediate state consistent with $f_0(980)$.
The di-pion invariant mass spectra for the pions in the decay $Y(4350)\to\psi'
\pi^+\pi^-$ could be consistent with a broad intermediate state with a mass 
around 600 MeV. This could be a $\sigma$ scalar meson \cite{sigma} and, in 
this case, one would not expect strange quarks in the current in 
Eq.~(\ref{field}). 

\begin{figure}[h]
\centerline{\epsfig{figure=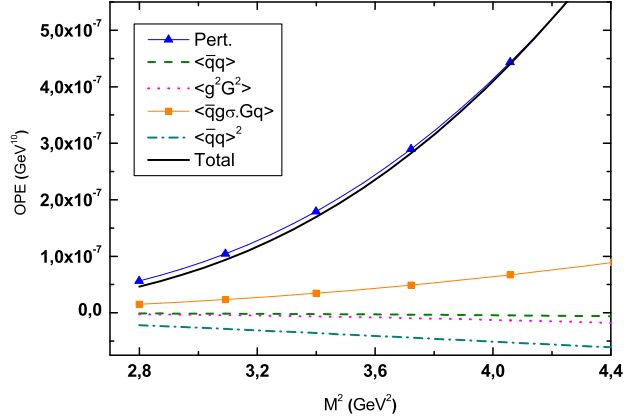,height=70mm}}
\caption{The OPE convergence for the sum rule with $m_s=m_q$, using 
$\sqrt{s_0} =4.9$ GeV. The solid with triangles, dashed, dotted, solid with 
squares, dot-dashed and solid lines give, respectively, the perturbative, 
quark condensate, gluon condensate, mixed condensate, four-quark condensate 
and total contributions.}
\label{figope43}
\end{figure}

Replacing the strange quarks in Eq.(\ref{field}) by the generic light quark 
$q$
and using $m_q=7.0\,\MeV$, we show, in Fig~\ref{figope43}, the OPE convergence
using $\sqrt{s_0}=4.9~\GeV$

Since now the quark condensate contribution is multiplyed by the light quark 
mass, its cotribution is now neglegible. From this figure we 
see that we get a good OPE convergence for $M^2\geq3.2~\GeV^2$.

Again the upper limits for $M^2$ are obtained by imposing that the pole 
contribution should be bigger than the continuum contribution. These values 
are given in Table I  for each value of $\sqrt{s_0}$.

\begin{center}
\small{{\bf Table I:} Upper limits in the Borel window for the sum rule with 
$m_s=m_q$.}
\\
\begin{tabular}{|c|c|}  \hline
$\sqrt{s_0}~(\GeV)$ & $M^2_{max}(\GeV^2)$  \\
\hline
 4.8 & 3.5 \\
\hline
4.9 & 3.8 \\
\hline
5.0 & 4.0 \\
\hline
\end{tabular}\end{center}

\begin{figure}[h]
\centerline{\epsfig{figure=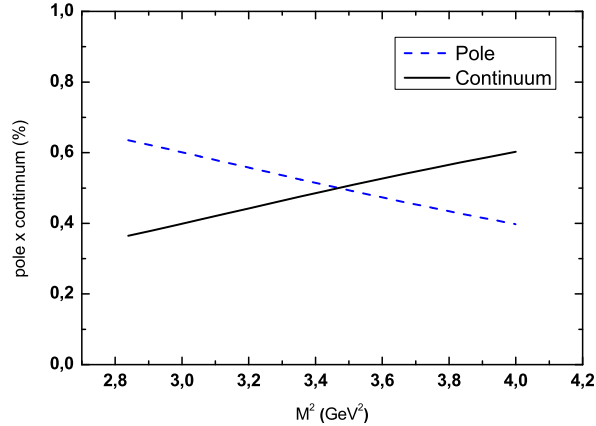,height=70mm}}
\caption{Same as Fig.~2 for the sum rule with $m_s=m_q$ and $\sqrt{s_0}=4.9~
\GeV$.}
\label{figpvczs}
\end{figure}

In Fig.~\ref{figpvczs} we show the relative continumm (solid line) versus 
pole (dashed line) contribution,  using $\sqrt{s_0}=4.9~\GeV$, from 
where we clearly see that the pole contribution is bigger than the continuum 
contribution for $M^2<3.5~\GeV^2$.

\begin{figure}[h]
\centerline{\epsfig{figure=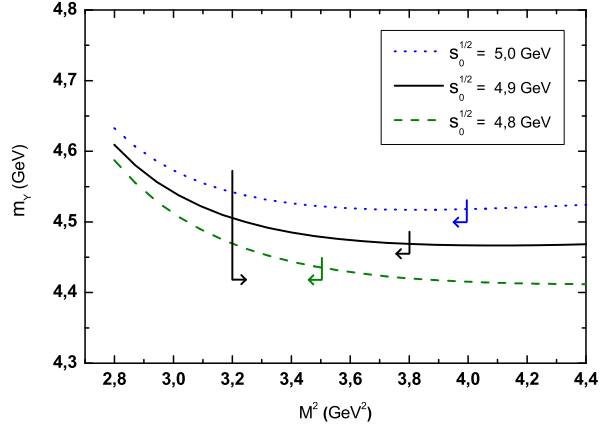,height=70mm}}
\caption{The light $m_Y$ meson mass as a function of the sum rule parameter
($M^2$) for different values of $\sqrt{s_0}$. The arrows delimit the regions 
allowed for the  sum rule.}
\label{figmzs}
\end{figure}

In this case the stability for the $m_Y$ mass is not as good as in the 
previous case, but it is still acceptable, in the allowed sum rule window, as 
a function of $M^2$ as can be seen by  Fig.~\ref{figmzs}.

Taking into account the variations on $M^2$, $s_0$, $\qq$ and $m_c$ in
the regions indicated above we get:
\beq
\lb{massy43}
m_Y=  (4.49\pm0.11)~\GeV~,
\enq
which is bigger than the $Y(4350)$ mass, but it is consistent with it 
considering the uncertainty.

\section{$D_{s0}\bar{D}_s^*$ molecule}

As pointed out in ref.~\cite{mol}, if the $X(3872)$,
and $Z^+(4430)$ are really molecular states, then many other molecules should 
exist. In particular a $D_{s0}(2317)\bar{D}_s^*(2110)$ molecule with $J^{PC}=
1^{--}$, could also decay into $\psi^\prime\pi^+\pi^-$ with a dipion mass 
spectra consistent with $f_0(980)$. Therefore, in this section we consider
a meson-meson current to check if this current could also describe the
meson $Y(4660)$. A current with $J^{PC}=1^{--}$ and a symmetrical combination 
between the scalar and vector mesons is given by:
\beq
j_\mu={1\over\sqrt{2}}[(\bar{s}_a\gamma_\mu c_a)(\bar{c}_b{s}_b)+(\bar{c}_a
\gamma_\mu s_a)(\bar{s}_b{c}_b)]\;.
\label{mol}
\enq
With this current we get for the spectral density, results very similar to 
the ones in Eq.(\ref{rhoope}), the only differences being the color factors. 
We get
a factor 3/4 for the diagrams where there is no gluons being exchanged between
the quark lines, and a factor 3/2 for the diagrams with gluons being exchanged
between the quark lines:
\beqa\label{opemol}
&&\rho^{pert}(s)=-{1\over 2^{10} \pi^6s}\int\limits_{\almi}^{\alma}
{d\al\over\alpha^3}
\int\limits_{\bemi}^{1-\al}{d\be\over\be^3}(1-\al-\be)
\left[(\al+\be)m_c^2-\al\be s\right]^3\left[m_c^2-2m_c^2(\alpha+\beta)+\al\be 
s\right],
\nn\\
&&\rho^{\qqs}(s)=-{9m_sm_c^2\qqs\over2^{6} \pi^4s}\int\limits_{\almi}^{\alma}
{d\al\over\alpha}
\int\limits_{\bemi}^{1-\al}{d\be\over\be}\left[(\al+\be)m_c^2-\al\be s\right],
\nn\\
&&\rho^{\lag G^2\rag}(s)={m_c^2\Gd\over2^{11}3\pi^6s}
\int\limits_{\almi}^{\alma} {d\al}
\int\limits_{\bemi}^{1-\al}{d\be\over\beta^3}(1-\al-\be)\left[4(2\al+2\be-1)
m_c^2-{3m_c^2\beta\over\alpha}(1+\al+\be)-\be s(7\alpha-3\be-3)\right],
\nn\\
&&\rho_1^{mix}(s)={m_c\mixs\over2^{7} \pi^4s}\int\limits_{\almi}^{\alma}
{d\al\over\alpha}
\int\limits_{\bemi}^{1-\al}{d\be\over\be^2}(2\alpha+\beta)\left[(\al+\be)m_c^2
-\al\be s\right],
\nn\\
&&\rho_2^{mix}(s)=-{m_s\mixs\over2^{7} \pi^4s}\int\limits_{\almi}^{\alma}
{d\al}\left[5m_c^2+\alpha^2s-{m_c^2\over1-\alpha}
-5m_c^2\int\limits_{\bemi}^{1-\al}{d\be\over\be}\right],
\nn\\
&&\rho^{\qqs^2}(s)=-{\qqs^2\over 2^43\pi^2}\left({5m_c^2\over s}-{1\over2}
\right)\sqrt{1-4m_c^2/s}.
\enqa

\begin{figure}[h]
\centerline{\epsfig{figure=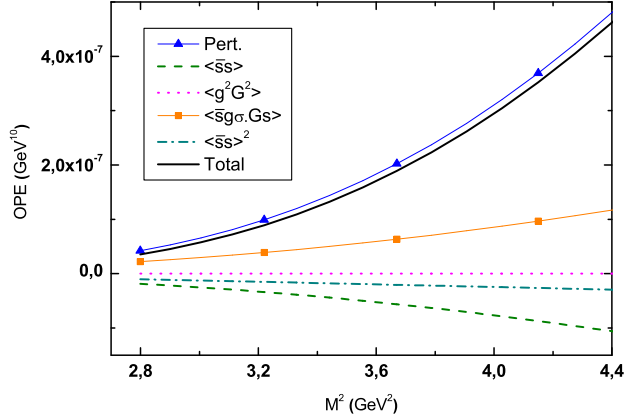,height=70mm}}
\caption{ The OPE convergence for the sum rule for the  $D_{s0}\bar{D}_s^*$ 
molecule with $\sqrt{s_0} = 4.9$ GeV.}
\label{convmol}
\end{figure}
With these changes, the mixed condensate is now, for the current in 
Eq.~(\ref{mol}),
more important than it was for the current Eq.~(\ref{field}), as can be seen 
in
Fig.~\ref{convmol}. For this current we find that the appropriate continuum
threshold range is $4.8\leq\sqrt{s_0}\leq5.0\GeV$ and that the OPE convergence
is  very good for $M^2\geq3.2~\GeV^2$. The upper limits for $M^2$ are 
are given in Table II  for each value of $\sqrt{s_0}$.

\eject
\begin{center}
\small{{\bf Table II:} Upper limits in the Borel window for the sum rule for
the $D_{s0}\bar{D}_s^*$ molecule.}
\\
\begin{tabular}{|c|c|}  \hline
$\sqrt{s_0}~(\GeV)$ & $M^2_{max}(\GeV^2)$  \\
\hline
 4.8 & 3.5 \\
\hline
4.9 & 3.7 \\
\hline
5.0 & 4.0 \\
\hline
\end{tabular}\end{center}

\begin{figure}[h]
\centerline{\epsfig{figure=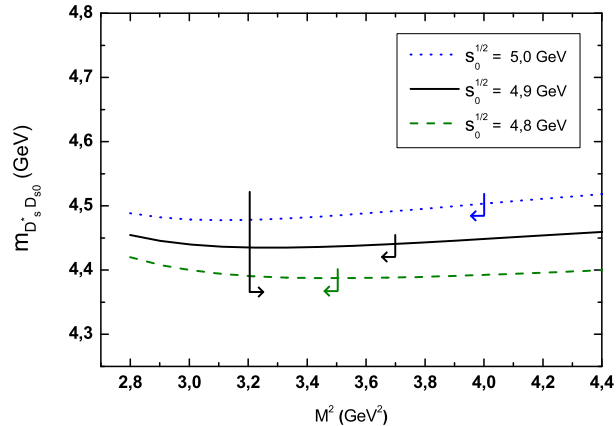,height=70mm}}
\caption{The ${D_{s0}\bar{D}_s^*}$ molecule mass as a function of the sum
rule parameter ($M^2$) for different values of $\sqrt{s_0}$. The arrows 
delimit the regions allowed for the  sum rule.}
\label{mmol}
\end{figure}

From  Fig.~\ref{mmol} we see that the Borel stability, as a function of $M^2$ 
is very good, in the allowed sum rule window.

Taking into account the variations on $M^2$, $s_0$, $\qq$ $m_s$ and $m_c$ in
the regions indicated above we get:
\beq
\lb{massmols}
m_{D_{s0}\bar{D}_s^*}=  (4.42\pm0.10)~\GeV~,
\enq
which is more in agreement with the $Y(4350)$ mass than with the $Y(4660)$
mass. It is important to mention that even if we use $\sqrt{s_0}=5.1~
\GeV$, which is the central value used for $Y(4660)$, 
we get $m_{D_{s0}\bar{D}_s^*}\sim
4.5~\GeV$, still in agreement, considering the errors, with the value in
Eq.~(\ref{massmols}). Therefore, we have to conclude that the $Y(4660)$ meson
is better explained with the diquark-antidiquark current in Eq.~(\ref{field})
than with the molecular current in Eq.~(\ref{mol}). To conclude if we
can associate this molecular state with the meson $Y(4350)$ we need a better
understanding of the di-pion invariant mass spectra for the pions in the 
decay $Y(4350)\to\psi'\pi^+\pi^-$. From the spectra given in 
ref.~\cite{babarconf}, it seems to us that the $Y(4350)$ is more consistent
with a non-strange four-quark state than with a ${D_{s0}\bar{D}_s^*}$ 
molecular state. However, from the mass obtained from the sum rules we
see that we can explain the $Y(4350)$ meson better as a ${D_{s0}\bar{D}_s^*}$ 
molecular state.

It is also important to notice that the central mass obtained for the 
${D_{s0}\bar{D}_s^*}$ molecule is very close to the ${D_{s0}(2317)
\bar{D}_s^*(2110)}$ threshold, indicating that this channel might be 
forbidden in the $Y(4350)$ decay.

\section{$D_{0}D^*$ molecule}

Similarly to what was done in section IV, we can consider a scalar-vector 
charmed mesons molecule: $D_{0}\bar{D}^*$ with $J^{PC}=1^{--}$. For this we 
have only to change the strange quarks in Eq.(\ref{mol}) by the generic light 
quark $q$. In this case we get a good OPE convergence for $M^2\geq 3.2~\GeV^2$
and the continuum threshold in the range $4.6\leq\sqrt{s_0}\leq4.8~\GeV$.
The upper limits for $M^2$ are are given in Table III  for each value of 
$\sqrt{s_0}$.

\begin{center}
\small{{\bf Table III:} Upper limits in the Borel window for the sum rule for
the $D_{0}\bar{D}^*$ molecule.}
\\
\begin{tabular}{|c|c|}  \hline
$\sqrt{s_0}~(\GeV)$ & $M^2_{max}(\GeV^2)$  \\
\hline
 4.6 & 3.3 \\
\hline
4.7 & 3.5 \\
\hline
4.8 & 3.7 \\
\hline
\end{tabular}\end{center}

\begin{figure}[h]
\centerline{\epsfig{figure=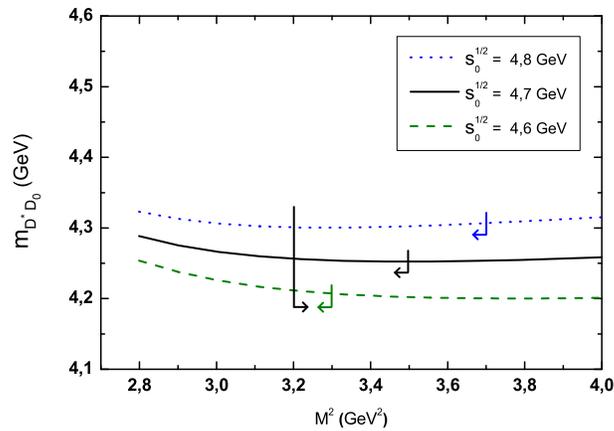,height=70mm}}
\caption{The ${D_{0}\bar{D}^*}$ molecule mass as a function of the sum rule 
parameter ($M^2$) for different values of $\sqrt{s_0}$. The arrows delimit 
the regions allowed for the  sum rule.}
\label{mmolq}
\end{figure}

Again we get a very good Borel stability as a function of $M^2$, in the 
allowed sum rule window, as can be seen by   Fig.~\ref{mmolq}. 
Considering the variations on $M^2$, $s_0$, $\qq$ and $m_c$ in
the regions indicated above we get:
\beq
\lb{massmol}
m_{D_{0}\bar{D}^*}=  (4.27\pm0.10)~\GeV~,
\enq
in excelent agreement with the mass of the meson $Y(4260)$.
Again, to conclude if we
can associate this molecular state with the meson $Y(4260)$ we need a better
understanding of the di-pion invariant mass spectra for the pions in the 
decay $Y(4260)\to J/\psi\pi^+\pi^-$. From the spectra given in 
ref.~\cite{babarconf}, it seems that the $Y(4260)$ is  consistent
with a non-strange molecular state ${D_{0}\bar{D}^*}$.

Regarding charmed non-strange scalar mesons, a broad enhancement ($D_0$) mass
distribution has been observed \cite{befo}. Its mass is now compiled 
as \cite{pdg} $m_{D_0}=2352\pm50~\MeV$. Therefore, the ${D_{0}\bar{D}^*}$
threshold is around 2360 MeV  and is 100 MeV above the molecule mass in 
Eq.~(\ref{massmol}).

\section{Conclusions}
In conclucion, we have presented a QCDSR analysis of the two-point
functions for tetraquark charmonium states with $J^{PC}=1^{--}$, to study 
the
charmonium $Y$ mesons recently observed by BaBar and BELLE Collaborations.
We have considered two kinds of currents: a diquark-antidiquark tetraquark 
state with a symmetric spin distribution $[cq]_{S=1}[\bar{c}\bar{q}]_{S=0}+
[cq]_{S=0}[\bar{c}\bar{q}]_{S=1}$ (where the generic quark $q$ can be either
a light $u$ or $d$ quark, or a strange, $s$, quark), and a molecular state
with a symmetrical combination between scalar and vector charmed mesons.

Our findings indicate that the $Y(4660)$ meson can be very well described
by a diquark-antidiquark $(cs)(\bar{c}\bar{s})$ tetraquark state.
This  quark content is consistent with the di-pion invariant mass spectra 
shown in ref.~\cite{babarconf}, which shows that there is some indication 
that the $Y(4660)$ has a well defined di-pion intermediate state 
consistent with $f_0(980)$.

In the case of the $Y(4350)$ meson, its mass can be reproduced, considering
the errors, if we assume that it is a diquark-antidiquark $(cq)(\bar{c}
\bar{q})$ tetraquark state, or a $D_{s0}\bar{D}_s^*$ molecule. 
Since an indication from its quark content can be obtained from
the di-pion invariant mass spectra in the decay $Y(4350)\to \psi'\pi^+\pi^-$,
we need a better understanding of it before we can reach a definite 
conclusion about its structure.

We also found that the $Y(4260)$ meson can be very well described
by a $D_{0}\bar{D}^*$ molecular state.

\section*{Acknowledgements}
{The authors would like to thank Stephan Narison for many discussions
and fruitful collaboration.
This work has been partly supported by FAPESP and CNPq-Brazil.}



\begin{references}

\bibitem{belle1} BELLE Coll.,  K. Abe  {\it et al.}, hep-ex/050719.

\bibitem{vol} M.B. Voloshin, arXiv:0711.4556 [hep-ph].

\bibitem{belle2} BELLE Coll., S.-K. Choi  {\it et al.},
Phys. Rev. Lett. {\bf 91}, 262001 (2003).

\bibitem{belle3} BELLE Coll., K. Abe  {\it et al.}, arXiv:0708.1790 [hep-ex].

\bibitem{babar1} BaBar Coll.,  B. Aubert  {\it et al.}, Phys. Rev. Lett.
{\bf95}, 142001 (2005).

\bibitem{babar2} BaBar Coll.,  B. Aubert  {\it et al.}, Phys. Rev. Lett.
{\bf98}, 212001 (2007).

\bibitem{belle4} BELLE Coll., X.L. Wang  {\it et al.}, Phys. Rev. Lett.
{\bf99}, 142002 (2007).

\bibitem{zhu} S.-L. Zhu, arXiv:0707.2623 [hep-ph]; hep-ph/0703225;
K.K. Seth, arXiv:0712.0340 [hep-ex]

\bibitem{babarconf} R. Faccini, arXiv:0801.2679 [hep-ex].

\bibitem{ghm} F.-K. Guo, C. Hanhart and U.-G. Meissner, arXiv:0803.1392
 [hep-ph].

\bibitem{qiao} C.F. Qiao, arXiv:0709.4066 [hep-ph].

\bibitem{dzy} G.-J. Ding, J.-J. Zhu and M.-L. Yan, arXiv:0708.3712 [hep-ph].

\bibitem{maiani} L. Maiani, V. Riquer, F. Piccinini and A.D. Polosa, Phy. Rev.
{\bf D72}, 031502 (2005).

\bibitem{gerry} G.A. Miller, Phys. Rev. {\bf C70}, 022202 (2004).

\bibitem{svz} M.A. Shifman, A.I. and Vainshtein and V.I. Zakharov,
Nucl. Phys. {\bf B147}, 385 (1979).

\bibitem{rry} L.J. Reinders, H. Rubinstein and S. Yazaki, Phys. Rept.
{\bf 127}, 1 (1985).

\bibitem{SNB} For a review and references to original works, see
e.g., S.
Narison, {\it QCD as a theory of hadrons,
Cambridge Monogr. Part. Phys. Nucl. Phys. Cosmol.} {\bf 17}, 1 (2002)
[hep-h/0205006]; {\it QCD
spectral sum rules ,  World Sci. Lect. Notes Phys.} {\bf 26}, 1 (1989);
{ Acta Phys. Pol.} {\bf B26}, 687 (1995); { Riv. Nuov. Cim.} {\bf 10N2}, 1
(1987); { Phys. Rept.} {\bf 84}, 263 (1982). 

\bibitem{x3872} R.D. Matheus, S. Narison, M. Nielsen and J.-M. Richard,
 Phys. Rev. {\bf D75}, 014005 (2007).

\bibitem{nos} S.H. Lee, A. Mihara, F.S. Navarra and M. Nielsen, Phys. Lett. 
{\bf B661}, 28 (2008).

\bibitem{io1} B.~L. Ioffe, Nucl.\ Phys.\ {\bf B188}, 317 (1981);
           {\bf B191}, 591(E) (1981).

\bibitem{BAGAN} Bagan et al., {Nucl. Phys.} {\bf B254} (1985) 55; D.J. 
Broadhurst and S. Generalis, {Phys. lett.} {\bf B139}  (1984) 85.

\bibitem{hen} K.-C. Yang, W.-Y.P. Hwang, E.M. Henley and L.S. Kisslinger, 
Phys. Rev. {\bf D47}, 3001 (1993).

\bibitem{tcc}   F.S.~Navarra, M.~Nielsen and S.H. Lee, Phys. Lett. 
{\bf B649}, 166 (2007).

\bibitem{narpdg} S. Narison, Phys. Lett. {\bf B466}, 345 (1999);
S. Narison,  Phys. Lett. {\bf B361}, 121 (1995);
S. Narison, Phys. Lett. {\bf B387}, 162 (1996). S. Narison,  Phys.  Lett.
{\bf B624}, 223 (2005).

\bibitem{sigma} E.M. Aitala {\it et al},  Phys. Rev. Lett. {\bf 86}, 770 
(2001). 


\bibitem{mol} S.H. Lee, M. Nielsen, U. Wiedner, arXiv:0803.1168 [hep-ph].

\bibitem{befo} BELLE Coll.,  K. Abe  {\it et al.}, Phys. Rev. {\bf D69},
112002 (2004).

\bibitem{pdg} Particle Data Group, S. Eidelman {\it et al}, J. Phys.
{\bf G33}, 1 (2006).

\end{references}
\end{document}